\documentclass[draft]{agujournal2019}
\usepackage{url} 
\usepackage{lineno}
\usepackage[inline]{trackchanges} 
\usepackage{soul}
\draftfalse

\journalname{Geophysical Research Letters}

\begin{document}

%
%


\title{A Turbulent Heating Model Combining Diffusion and Advection Effects for Giant Planet Magnetospheres}

%
%




\authors{C. S. Ng\affil{1}, B. R. Neupane\affil{1}, P. A. Delamere\affil{1}, P. A. Damiano\affil{1}}

\affiliation{1}{Geophysical Institute, University of Alaska Fairbanks, Fairbanks, AK, USA}





\correspondingauthor{C. S. Ng}{cng2@alaska.edu}




\begin{keypoints}
\item A new and improved model for the heating of the magnetospheres of Jupiter and Saturn by magnetohydrodynamic turbulence is developed
\item The model combines effects from diffusion and advection such that each is dominant when the radial flow velocity is small or large
\item Predictions of the temperature and radial flow velocity profiles agree better with Jupiter and Saturn observations than previous models
\end{keypoints}

%
%

%
%


\begin{abstract}
The ion temperature of the magnetospheres of Jupiter and Saturn was observed to increase substantially from about 10 to 30 planet radii. 
Different heating mechanisms have been proposed to explain such observations, 
  including a heating model for Jupiter based on MHD turbulence with flux-tube diffusion. 
More recently, 
  an MHD turbulent heating model based on advection was shown to also explain the temperature increase at Jupiter and Saturn. 
We further develop this turbulent heating model by combining effects from both diffusion and advection. 
The combined model resolves the physical consistency requirement that diffusion should dominate over advection 
  when the radial flow velocity is small and vice versa when it is large. 
Comparisons with observations show that previous agreements,
   using the advection only model,
   are still valid for larger radial distance. 
Moreover, 
  the additional heating by diffusion  
  results in a better agreement with the temperature 
  observations for smaller radial distance.
\end{abstract}

\section*{Plain Language Summary}
The ion temperature of the magnetospheres of Jupiter and Saturn was observed to increase substantially near the planet. 
This suggests that there should be some heating sources to counter the cooling effect due to expansion. 
There have been several models trying to explain such observation using different heating mechanisms, 
  including a heating model for Jupiter based on turbulence and diffusion effects,
  as well as a model based on advection effects for Jupiter and Saturn. 
We further develop a heating model by combining effects from both diffusion and advection. 
The combined model resolves the physical consistency requirement that diffusion should be stronger than advection 
  nearer to the planet, but shifting to the opposite farther away. 
Comparisons with observations show that previous agreements using the advection only model 
  are still valid,
  and are improved by including diffusion nearer to the planet.

%
%

%


%
%
%
%

\section{Introduction}
One major unresolved problem in the physics of giant magnetospheres 
  is to find the mechanism responsible for the observed increase of 
  ion temperature with radial distance 
  \cite <e.g.,>{Bagenal11}.
Such observations show that the ion temperature of the magnetospheres of Jupiter and Saturn 
  increase substantially, about a factor of 3, from about 10 to 30 planet radii. 
This suggests that there should be some heating sources to counter 
  the cooling effect due to adiabatic expansion. 
There have been several models proposed to explain 
  such observations using different heating mechanisms 
  \cite <e.g.,>{Hill83_Dessler}. 
\citeA{Saur04} considered the possibility of turbulent heating of
  Jupiter's magnetodisc (10 to 40 ${\rm R_J}$) based on magnetohydrodynamic (MHD) turbulence,
  and flux tube diffusion.
Another turbulent heating model for Jupiter was developed by 
  \citeA{Ng-etal-JGR-2018},
  in which the outward plasma transport is dominated by advection rather than diffusion,
  as inspired by turbulent heating models of the solar wind 
  \cite <e.g.,>{Ng10a}.
They found that the observed heating rate density due to MHD turbulence can provide 
  enough heating to explain the observed increase of ion temperature to about 30 ${\rm R_J}$.
More recently the same turbulence heating model was also 
  applied to Saturn's magnetosphere by 
  \citeA{https://doi.org/10.1029/2020JA027986},
  who also found a consistent relationship 
  between the heating rate density and the temperature increase.
  
The justification for the advection dominated turbulent heating model 
  was argued in details by \citeA{Ng-etal-JGR-2018},
  based on the observed fact that beyond about 10 ${\rm R_J}$ 
  the radial transport time becomes much shorter,
  or the radial outflow speed becomes much larger 
  \cite{Delamere10, Bagenal11, Bagenal16}.
While this model does give the required heating rate at the radial positions 
  up to about 30 planet radii for both Jupiter and Saturn,
  strictly speaking it cannot be  applied to positions near and inward of 10 planet radii 
  due to the fact the the radial outflow speed is still small there.
Moreover, 
  the temperature increases predicted by this model do seem to be significantly
  smaller than the observed data for both Jupiter and Saturn in the near planet positions
  \cite{Ng-etal-JGR-2018, https://doi.org/10.1029/2020JA027986}.

In this paper,
  we develop a new turbulent heating model by combining 
  the diffusion model of 
  \citeA{Saur04},
  and the advection model of 
  \citeA{Ng-etal-JGR-2018}.
The formulation for this model will be given in the next Section.
In Section~\ref{compare} we will apply this new model to both
  Jupiter and Saturn cases to compare the predicted 
  temperature, 
  as well as radial outflow speed, with observations.
Discussion and conclusion will be given in Section~\ref{conclusion}.
  
\section{A Combined Turbulent Heating Model}
  \label{model}
  
We start the derivation of the one-dimensional steady-state turbulent heating model
  by writing down the transport equations for mass,
  
\begin{equation}
  \dot{M} = 2\pi LR H m_i nV_r  - 2\pi m_i \frac{D_{LL}}{L^2} \frac{d}{dL}(nHL^3) \; .  
  \label{mdot}
\end{equation}  
In this equation, $\dot{M}$ is the mass rate to be transported out of the inner magnetosphere
  injected by Io of Jupiter or Enceladus of Saturn.
$R$ is the radius of the planet.
$L=r/R$ is the normalized radial position.
$H$ is the scale height of an equatorially confined plasma sheet.
$m_i$ is the average mass of ions to be transported out.
$n$ is the volume ion number density.  
$V_r$ is the radial outflow velocity.
$D_{LL}$ is the diffusion coefficient of the flux tube content of ions.
Similarly,
  the transport equations for heat is of the form
\begin{equation}
  q = \frac{3}{2LR}n^{5/3}V_r \frac{d}{dL}\left(pn^{-5/3}\right)
    -  \frac{3}{2HR^2 L^3} \frac{d}{dL}
    \left[ \frac{D_{LL}}{L^2} \frac{d}{dL} (nHk_B TL^5) \right] \; ,  
  \label{q}
\end{equation}  
where $k_B$ is the Boltzmann constant,
  $T$ is the ion temperature,
  $p = nk_B T$ is the ion pressure,
  $q$ is the heating rate density,
  assumed to be due to the dissipation of MHD turbulence in this paper.
On the right hand sides of Eqs.~(\ref{mdot}) and (\ref{q}),
  the first term is due to advection and 
  the second term is due to diffusion.
The previous turbulent heating models by 
  \citeA{Saur04} 
  and \citeA{Ng-etal-JGR-2018} can be recovered 
  by keeping only the diffusion terms or the advection terms respectively.

While the scale height $H$ as a function of $L$ can be 
  expressed as an empirical formula fitted to observations 
  \cite{Bagenal11, Bagenal16, Thomsen10, doi:10.1002/2017JA024117},
  for simplicity we can approximate it as either a constant 
  such as $H = 2R$ used by 
  \citeA{Ng-etal-JGR-2018},
  or a linear function such as $H = \theta L R$, 
  with the constant $\theta = 0.27$,
  used by 
  \citeA{https://doi.org/10.1029/2020JA027986}.
We will adopt the latter form in this paper for uniformity,
  although it is straightforward to derive alternate equations based on other
  models for the scale height.
The ion density is also assumed to have a power law profile of $n = n_0 L^{-\beta}$.
Following 
  \citeA{Saur04}, 
  we will also assume the diffusion coefficient is having a form of power law dependency 
  $D_{LL} = D_0 L^b$ such that the radial outflow velocity 
  can be solved from Eq.~(\ref{mdot}) as
\begin{equation}
  V_r = \left[\frac{\dot{M}}{2\pi R m_i} - (\beta - 4)D_0 n_0 \theta L^{b-\beta + 1}\right]
  \frac{1}{\theta RnL^2}  \; .  
  \label{vr}
\end{equation}  
For a diffusion only model without radial outflow,
  we then have $b = \beta -1$ and
\begin{equation}
  D_0 = \frac{\dot{M}}{2\pi R m_i (\beta - 4) n_0 \theta}  \; .  
  \label{d0}
\end{equation}  
Note that $b = \beta$ instead in 
  \citeA{Saur04},
  which is simply due to assuming a constant $H$,
  rather than the linear form used in this derivation.
If $b = \beta -1$ is a constant and with Eq.~(\ref{d0}),
  $V_r$ would be identically zero for all $L$,
  which is in contradiction with the 
  observed growth along $L$ 
  \cite <e.g.,>{Bagenal11}.
While such a growth in $V_r$ must be due to a dynamical mechanism,
  here we will simply model the growth by 
  assuming $b$ starts decreasing with $L$ from $\beta -1$ 
  (after a certain initial position $L = L_0$) and 
  compare with observations.

Substituting Eq.~(\ref{vr}) into Eq.~(\ref{q}),
  a second order ordinary differential equation for $T$ can be obtained as
\begin{equation}
  \frac{d^2 T}{dL^2} - \frac{\eta L^{\beta_1} + \beta_1 - 5}{L}
  \frac{d T}{dL} - \left[ \frac{2\beta}{3}\left(\eta L^{\beta_1} - \beta + 4\right)
  + (6 - \beta)(\beta_1 - 2) \right]\frac{T}{L^2} = - \xi  \; , 
  \label{T-equation}
\end{equation}  
where $\beta_1 = \beta - b - 1$,
\begin{equation}
  \eta = \frac{\dot{M}}{2\pi R m_i n_0 D_0 \theta} = \beta - 4  \; ,
  \label{eta}
\end{equation}  
and 
\begin{equation}
  \xi = \frac{2R^2 q L^{\beta - b}}{3 k_B n_0 D_0}   \; .
  \label{xi}
\end{equation}  

\begin{figure}
  \noindent\includegraphics[width=\textwidth]{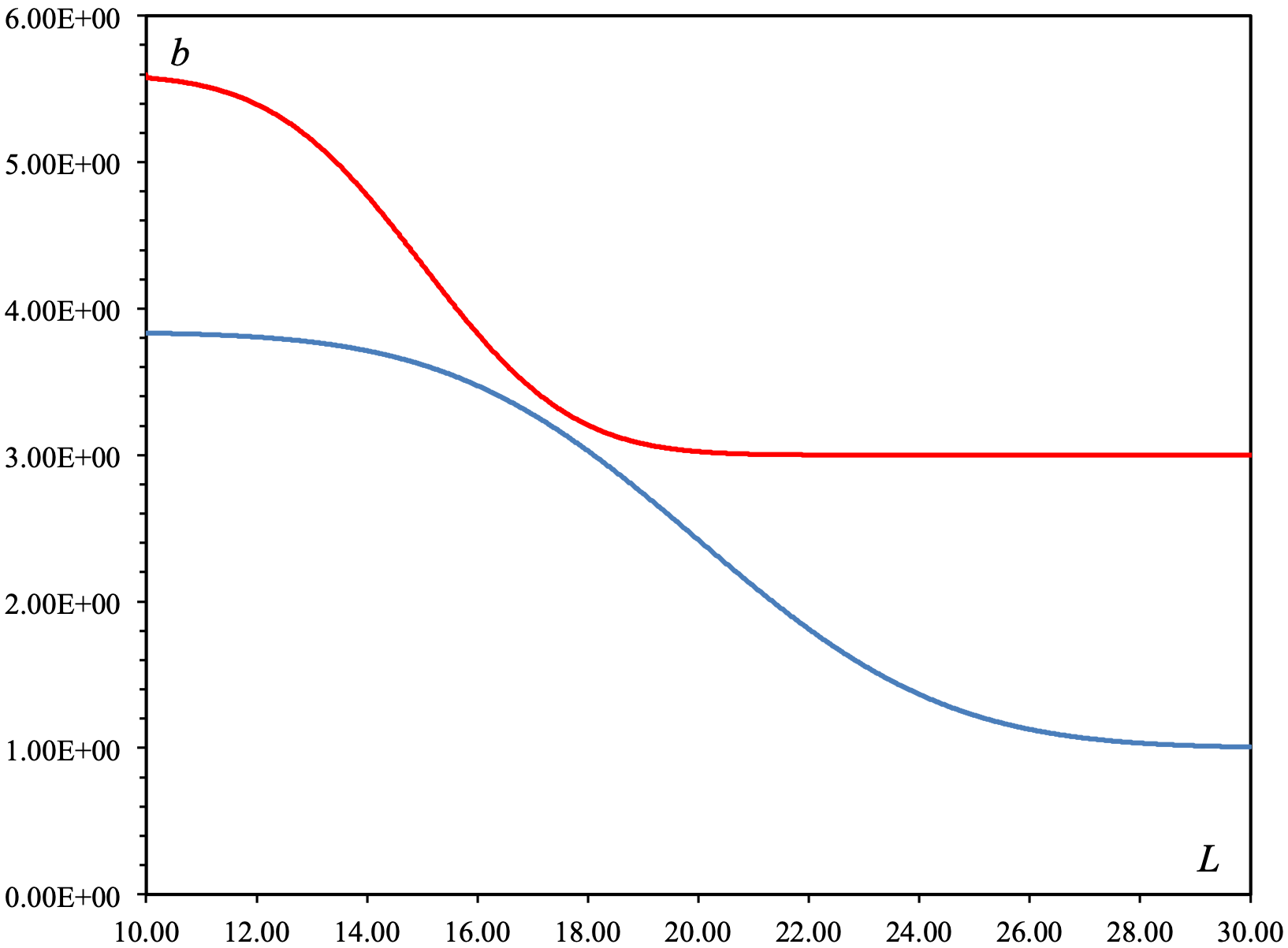}
\caption{The profile of the power law index $b$ as a function of $L$ for the Jupiter case
  (red curve), and the Saturn case (blue curve). }
\label{b}
\end{figure}

Since $b$ will be chosen to be less than $\beta -1$
  for larger $L$,
  $\beta_1$ is positive when $V_r$
  becomes larger.
In this case,
  terms proportional to $L^{\beta_1}$ dominate the left
  hand side of Eq.~(\ref{T-equation}) such that
  the advection only model by 
  \citeA{Ng-etal-JGR-2018}
  is recovered when $L$ is larger.
Eq.~(\ref{T-equation}) can be solved as a two-point 
  boundary value problem with
  $T(L_0) = T_0$ set for the small $L$ side.
The boundary condition for the large $L$ side is set
  by requiring that advection terms are larger than 
  diffusion terms on the left hand side of Eq.~(\ref{T-equation}).
Numerically this is imposed by choosing $T$ on the 
  large $L$ boundary such that there is no artificial
  oscillations in $L$ to make the $d^2T/dL^2$ term large.
From the experience in actually solving this equation numerically,
  this can be easily achieved.
Much effort has been spent by 
  \citeA{Kaminker17},
  \citeA{Ng-etal-JGR-2018} and
  \citeA{https://doi.org/10.1029/2020JA027986} 
  to determine 
  the heating rate density $q$ 
  by analyzing MHD turbulence spectra from magnetometer data
  for both Saturn and Jupiter.
Please refer to these papers for the data analysis methods,
  as well as various plots showing the distributions of $q$ 
  as functions of positions.
While the resulting $q$ values scatter over some ranges for different $L$,
  for the purpose of entering $q$ into the model calculations,
  it can be approximated by a fitting function
  of the form $q = q_0 L^s$,
  as obtained by 
  \citeA{Ng-etal-JGR-2018} and
  \citeA{https://doi.org/10.1029/2020JA027986}.

\begin{figure}
  \noindent\includegraphics[width=\textwidth]{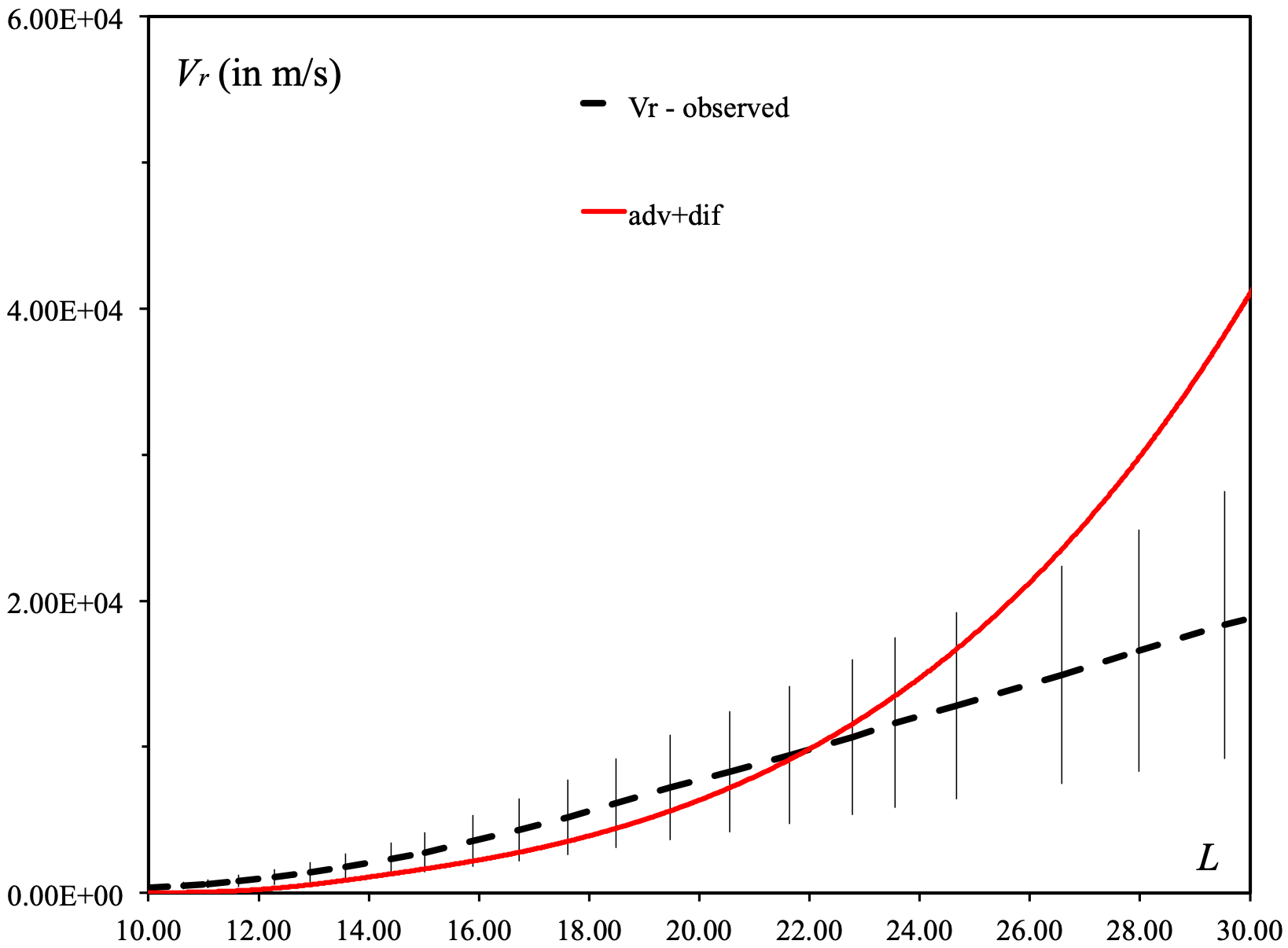}
\caption{Red curve: model output of the radial outflow velocity $V_r$ 
  as a function of $L$ for the Jupiter case. 
  Black dashed curve: observed $V_r$ from
   \citeA{Bagenal11} with $\pm 50\%$ vertical error bars.}
\label{J-vr}
\end{figure}

\section{Comparing Model Predictions with Observations}
  \label{compare}
We now apply the new turbulent heating model with both advection 
  and diffusion physics to the Jupiter case using data and parameters
  used by 
  \citeA{Ng-etal-JGR-2018}.
Let us first list the values of parameters we use:
  $n_0 = 1.9 \times 10^{14} {\rm m}^{-3}$,
  $\beta = 6.6$,
  $R = R_{\rm J} = 7 \times 10^7 {\rm m}$,
  $L_0 = 10$,
  $\dot{M} = 330 {\rm kg}/{\rm s}$,
  $m_i = 3.2 \times 10^{-26} {\rm kg}$,
  $\theta = 0.27$,
  $q_0 = 1.2 \times 10^{-14} {\rm W}/{\rm m}^3$,
  $s = -0.57$,
  and $T_0 = 1.7 \times 10^6 {\rm K}$.
  
The values of $L_0$ and $T_0$ are chosen here to be consistent
  with our published work.
In principle, 
  those values should be determined by physical chemistry models
  for the inner magnetospheres
  \cite{Delamere05, Fleshman13}.  
By such considerations,
  as well as observations,
   \cite <e.g.,>{sittler2008},
   $L_0$ could be chosen slightly inwards to about 8
   for Jupiter, and about 7 for Saturn.
Such small adjustments should not affect the conclusion of this paper,
  but might be needed for future investigations.

While there is a considerable amount of freedom to choose the 
  profile of $b$ as a function of $L$,
  many small adjustments simply produce qualitatively similar results.
Therefore we only try a few different profiles and 
  report here just one reasonable case with $b(L)$ shown
  as the red curve in Fig.~\ref{b}.   
With this choice of $b$,
  the radial outflow velocity can then be calculated using 
  Eqs.~(\ref{vr}) and (\ref{d0}),
  shown as the red curve in Fig.~\ref{J-vr}.
We see that for this profile of $b$,
  $V_r$ is mostly within $\pm 50 \%$ of the observed values from
  \citeA{Bagenal11},
  shown as the black dashed curve,
  which are derived from Galileo data,
  interpolated for the specific $\dot{M}$ value
  used in this calculation.
  The model $V_r$ does get higher than the observed values
  as $L$ getting near 30 when the heating
  model is expected to start failing.
  
\begin{figure}
  \noindent\includegraphics[width=\textwidth]{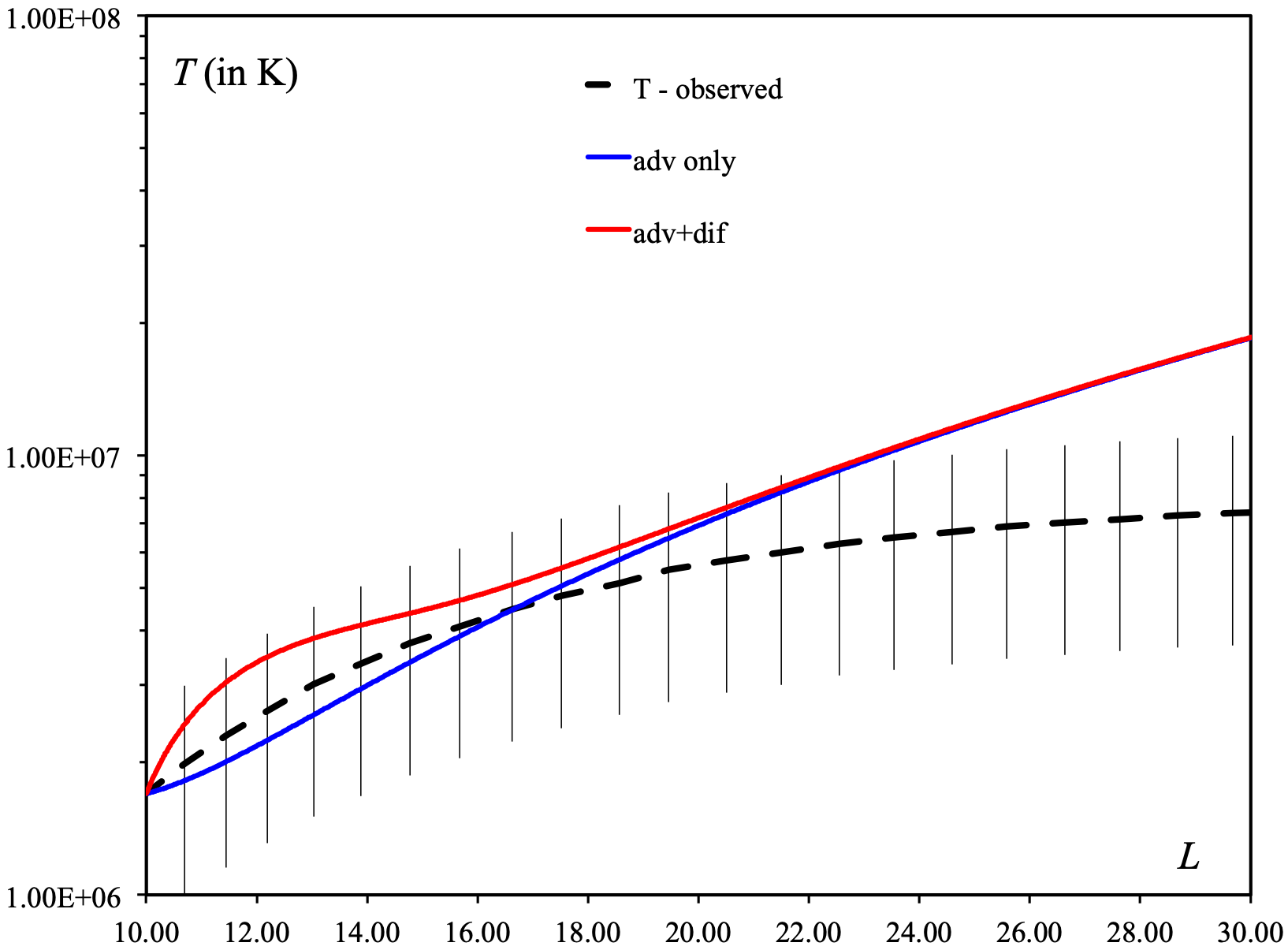}
\caption{Red curve: model output of the ion temperature $T$ 
  as a function of $L$ for the Jupiter case. 
  Blue curve: output of $T$ using the advection only model.
  Black dashed curve: observed $T$ from
   \citeA{Bagenal11} with $\pm 50\%$ vertical error bars.}
\label{J-T}
\end{figure}

 The corresponding ion temperature $T$ can then be calculated
  using Eq.~(\ref{T-equation}),
  shown as the red curve in Fig.~\ref{J-T}.
The blue curve in Fig.~\ref{J-T} shows $T$ calculated using
  the advection only model,
  i.e., keeping only terms proportional to $L^{\beta_1}$ on the left
  hand side of Eq.~(\ref{T-equation}).
We see that the output from the combined model shows a much
  faster increase from $T_0$ than the advection only model.
Then the two models actually give virtually the same $T$
  after $L \sim 20$.
Overall,
  the output from the advection only model is well within a
  factor of two of the output from the combined model.
This provides a justification that the advection only model
  used by 
  \citeA{Ng-etal-JGR-2018} 
  is valid,
  as long as the radial outflow velocity begins to grow larger.
Also shown as the black dashed curve in 
  Fig.~\ref{J-T} is the observed $T$,
  again from 
  \citeA{Bagenal11}.
We see that the predicted $T$ from both models actually 
  fall within $\pm 50 \%$ of the observed values for 
  over half of the range shown.
The predicted $T$ becomes consistently larger than
  the observations further in the $L$ range 
  but is still within a factor of three.

\begin{figure}
  \noindent\includegraphics[width=\textwidth]{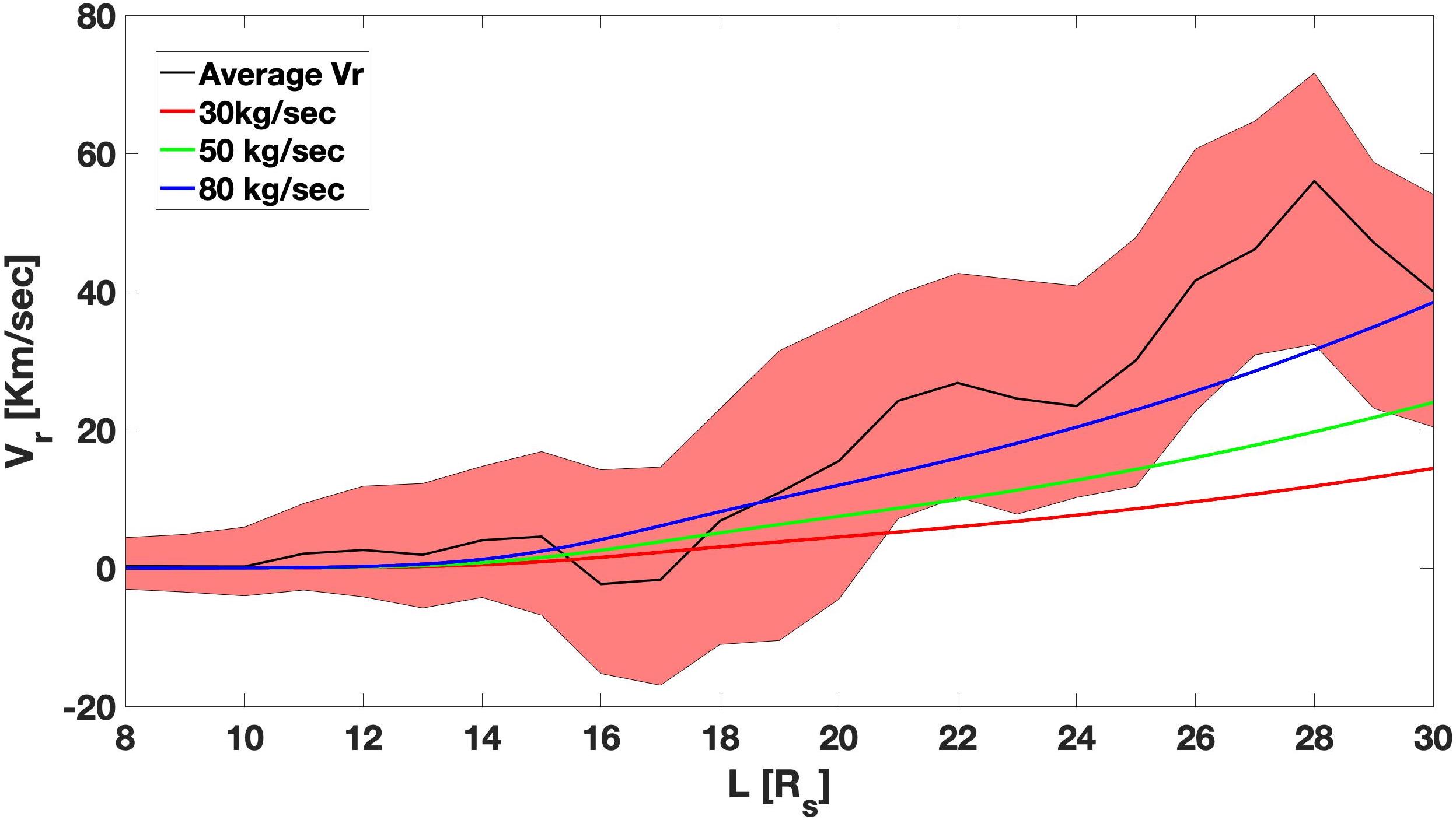}
\caption{Red, green, and blue curves: model outputs of the radial outflow velocity $V_r$ 
  as a function of $L$ for the Saturn case using 
  $\dot{M} $ = 30, 50, and 80 kg/s. 
  Black curve: observed $V_r$ from
   \citeA{doi:10.1002/2017JA024117} with the shaded
   area indicating values between 25th and 75th percentile.}
\label{S-vr}
\end{figure}

\begin{figure}
  \noindent\includegraphics[width=\textwidth]{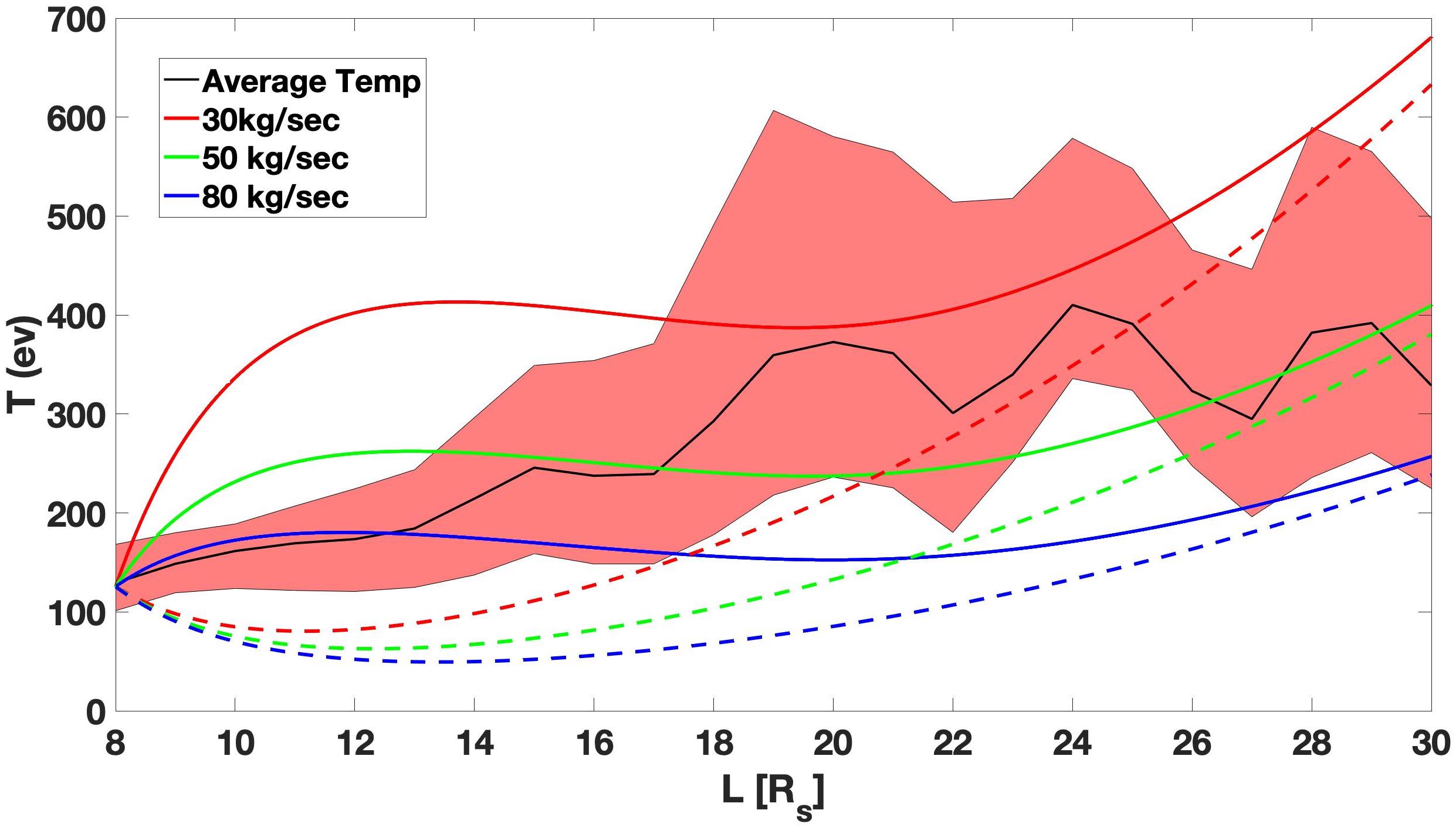}
\caption{Red, green, and blue curves: model outputs of the ion temperature $T$ 
  as a function of $L$ for the Saturn case using 
  $\dot{M}$ = 30, 50, and 80 kg/s,
  with solid curves from the combined model,
  and dashed curves from the advection only model.
  Black curve: observed $T$ from
   \citeA{doi:10.1002/2017JA024117} with the shaded
   area indicating values between 25th and 75th percentile.}
\label{S-T}
\end{figure}

Let us now apply the same model 
  to the Saturn case using data and parameters
  used by 
  \citeA{https://doi.org/10.1029/2020JA027986}.
Let us first list the values of parameters we use:
  $n_0 = 1.79 \times 10^{11} {\rm m}^{-3}$,
  $\beta = 4.84$,
  $R = R_{\rm S} = 6 \times 10^7 {\rm m}$,
  $L_0 = 8$,
  $m_i = 3 \times 10^{-26} {\rm kg}$,
  $\theta = 0.27$,
  $q_0 = 2.51 \times 10^{-16} {\rm m}^3 /{\rm W}$,
  $s = -0.3$,
  and $T_0 = 125.6 {\rm eV}$.
As in the Jupiter case,
  a $b$ profile is chosen after a few trials to give
  reasonable $V_r$ profiles,
  shown as the blue curve in Fig.~\ref{b}.
It has values quite different from the Jupiter case
  since the $\beta$ values are different.
Due to some uncertainty in the value of $\dot{M}$
  \cite{Delamere13b, Fleshman13},
  three values, 
  30, 50, and 80 kg/s,
  are used,
  as in 
  \citeA{https://doi.org/10.1029/2020JA027986}.
The $V_r$ profiles for these three cases 
  calculated from Eqs.~(\ref{vr}) and (\ref{d0}) are shown
  in Fig.~\ref{S-vr} as the red, green, and blue curves
  respectively.
We see that while the blue curve seems to be closest
  to the observed values (black curve) from
  \citeA{doi:10.1002/2017JA024117},
  the green curve is still mainly within the uncertainty region
  (the shaded
   area indicating values between 25th and 75th percentile).
Only the red curve is significantly below observations.

The corresponding ion temperature $T$ calculated
  from Eq.~(\ref{T-equation}) are
  shown in Fig.~\ref{S-T} for the same three $\dot{M}$
  cases as the red, green, and blue curves.
The observed temperature shown as the black curve
  is again from  
  \citeA{doi:10.1002/2017JA024117}.
We see that the green solid curve fits the observations the best,
  while the red solid curve is significantly higher,
  and the blue solid curve is significantly lower.
Combining both the $V_r$ and $T$ comparisons,
  the case for the green curves, 
  or $\dot{M} = 50 {\rm kg/s}$,
  seems to give an overall satisfactory predictions.
In Fig.~\ref{S-T} we have also plotted predictions
  using the advection only model as dashed
  curves,
  with the same color for each $\dot{M}$ case.  
Comparing with those curves,
  the combined model again gives a much faster increase of 
  $T$ starting from $L_0$,
  instead of having initial decreases,
  which now fits much better with observations,
  while the predictions at larger $L$ are essentially at
  the same levels.
Therefore the Saturn case again shows that the combined
  model indeed provides better agreement,
  and also recovers the advection only model for larger $L$.

\section{Conclusion}
   \label{conclusion}

In this paper,
  we have developed the formulation
  for a one-dimensional steady-state turbulent 
  heating model for the inner magnetospheres
  of giant planets,
  by combining both the diffusion
  and advection effects.
Combining these effects into a single model
  provides a better theoretical foundation than either the diffusion only 
  or advection only approaches of
  \citeA{Saur04},
  and 
  \citeA{Ng-etal-JGR-2018}
  respectively.
This is because the new model will consistently
  have the diffusion effects being dominant when
  the radial outflow speed $V_r$ is small,
  but will change to advection effects being
  dominant when $V_r$ is larger.
In practice,
  we also show that the combined model
  does give better comparisons with 
  observations of both $V_r$,
  and the ion temperature $T$,
  by repeating the studies done previously for 
  Jupiter by 
  \citeA{Ng-etal-JGR-2018},
  and for Saturn by 
  \citeA{https://doi.org/10.1029/2020JA027986}.
The new calculations from the combined model
  also provide a justification for the advection only model
  for the use in larger radial positions $L$
  when $V_r$ has increased substantially,
  since the two models give almost the same outputs
  for larger $L$.
With this new model and the comparisons with observations,
  there is more confidence that MHD turbulence
  can indeed provide enough heating to explain the increase
  with ion temperature in the inner magnetospheres of the
  giant planets.
  
While this turbulent heating model is now on firmer ground,
  it is still far from being a dynamical model,
  since we have not included an equation of motion that
  can provide a mechanism for the increase of $V_r$.
Rather,
  we have simply tuned the model to better compare
  with observed profiles of $V_r$ and $T$.
Such a study is obviously a worthwhile future research direction,
  but is outside the scope of this paper.
We additionally neglect other effects which contribute to the beginning 
  of the break down of the model  after about $L = 30$ such as
  the beginning of the breakdown of corotation,
  and the return of flux tubes,
  as well as other possible heat loss mechanisms.
Therefore,
  there are several potential directions that can be taken to facilitate 
  further improvement of the model.

\acknowledgments
This work is supported by NASA grant 80NSSC20K1279.

\noindent
Data Availability Statement

The new model outputs in this paper is based on parameters from 
  previous data analysis in the papers by 
  \citeA{Ng-etal-JGR-2018} and  \citeA{https://doi.org/10.1029/2020JA027986}.
  Please refer to these two papers for the availability of data used in those studies.


%
%


%
%
%
%
%

\end{document}